\documentstyle[12pt]{article}
\date{ }

\title{The thermodynamics of portfolios}
\author{E. W. Piotrowski\\ Institute of Theoretical Physics,
University of Bia\l ystok,\\ Lipowa 41, Pl 15424 Bia\l ystok,
Poland\\ e-mail: ep@alpha.uwb.edu.pl\\ and J. S\l adkowski
\\ Institute of Physics, University of Silesia, \\ Uniwersytecka
4, Pl 40007 Katowice, Poland \\ e-mail: sladk@us.edu.pl}

\begin{document}
\maketitle
\def\Z{{\bf Z\!\!Z}}
\begin{abstract}
We propose a new method of valuation of portfolios and their
respective investing strategies. To this end we define a canonical
ensemble of portfolios that allows to use the formalism
thermodynamics.

\end{abstract}



\section{Introduction}
Physicist, contrary to mathematicians, have only occasionally
investigated economic systems. Recently, however, a growing number
of papers of relevance to economics is being published in physics
journals and conferences proceedings \cite{1}. Moreover, H.
Markowitz encouraged outsiders to engage in research on economics
in his Nobel Prize Lecture \cite{2}: {\it "I believe that
microscopic market simulations have an important role to play in
economics and finance. If it takes people from outside economics
and finance -- perhaps physicists -- to demonstrate this role it
won't be for the first time that outsiders have made substantial
contribution to these fields"}. Physicists  have good command of
stochastic processes and statistical physics so it is hardly
surprising that they successfully use their minds to analyzing
economic systems. Other physical concepts  probably have analogue
in economics, e.g. such notion as gauge symmetry can be identified
in financial markets \cite{3}. Another important and exciting
branch of finances is the portfolio theory \cite{4}. Here various
physical concepts have direct analogies. Ultrametricity \cite{5}
commonly used in spin glasses can be used to describe distance
between stocks \cite{6}. Spin glasses seem to have lots in common
with portfolio theory \cite{7, 8}. We would like to propose a
method of valuation of portfolios and investing strategies that
stems from thermodynamics. Investors and portfolio managers set up
their portfolio according to the market information available and
their lore. They lose or gain. It is easy to gain when all prices
are raising and lose when everything is falling down. But which
part of the gain is due to skill and qualities of the investor?
Which part is simple a result of the actual state of the market?
Our model allows for a temperature like parameter that measures
the quality and professionalism of the investor.

\section{The portfolio description}

A portfolio is a package of various assets (shares, bonds,
derivative instruments, etc) that can be exchanged on the market.
If we denote by $a_{i},\ i= 0,1, \dots N$ the unit of the $i$-th
asset then the portfolio $P$ as $$ P=\sum_{i=0}^N \alpha _{i}
a_{i}, \eqno(1)$$ where $\alpha _{i} \in \Z $ is the number of
units of the $i$-th asset in the portfolio $P$. The coefficients
$\alpha _{i}$ can be negative because the stock exchange
regulations allow for selling assets that the portfolio owner do
not possesses (short selling). One usually supposes that one
asset, say $\alpha _{0}$, in $P$ can be exchanged with any other
asset at any time (money). An external observer describes the
moves performed by the portfolio manager as a draw in the
following  lottery. Let $p_{i}, i=1,...,N$ be the probability of
the purchase of the $w_{i}$ units of the $i$-th asset. The weights
$w_{i}$ are given before the lottery come into operation. The
value of the portfolio (return), $-c_{00\kappa}$, fulfills the
following balance equation at an arbitrary moment after the draw :
$$c_{00\kappa}\left( c_{0}, \dots ,c_{N}\right) + \sum_{k=0}^{N}
\left[ \kappa = k\right] c_{k} w_{k}=0, \eqno(2)$$ where $\kappa$
denotes the random variable taking values from {$0,\dots ,N$}, the
assets numbers. The expression $[sentence]$ takes value 0 or 1 if
the $sentence$ is false or true, respectively (Iverson convention)
\cite{9}. The coefficient $c_{i}$ denotes the present relative
price of a unit of the asset $a_{i}$,
$c_{i}=\frac{u_{i}}{\overline{u}_{i}}$ where $u_{i}$ is the
present price of the $i$-th asset and $\overline{u}_{i}$ its price
at the moment of drawing. At the moment of drawing the balance
equation takes the form: $$c_{00\kappa}\left( 1, \dots ,1\right) +
\sum_{k=0}^{N} \left[ \kappa = k\right] w_{k}=0, \eqno(3)$$ If
$\alpha_{0}$ represents the basic currency (money) used to define
the financial value of the portfolio then we have $c_{0}=1$ during
considered period. One can show that the expectation value of
$w_{0}\left[\kappa =0\right]$ is the Legendre transform of the
mean value of the portfolio.

\section{The canonical portfolio}

Let us consider a portfolio defined by the weights $w_{i}$ and the
uniform distribution of probabilities $p_{0}=p_{1}=\dots
=p_{N}=\frac{1}{N+1}$. This portfolio has the same expectation
value of the return $E(c_{00\kappa})$ as the uniform portfolio
with weights $\frac{w_{0}}{N+1} ,\dots , \frac{w_{N}}{N+1}$ (every
asset is included). It is interesting to find out how many of the
portfolio owners (managers) are in this situation. To this end we
consider two types of investors. The first group, called {\em
zombies}\/, consists of investors whose moves on the market are
fully deterministic. They react according to the market condition
distributing their capital into various assets or not and act in
the same way  in the same situations. The second group, called
{\em gamblers}, are indeterministic. Their moves can be different
in analogous market situations. This does not mean that gamblers
often change their strategies. We simply cannot predict their
moves because part of their knowledge and past experiences are not
available for us as external observers. Is there a qualitative way
of measuring information on investor's behaviour? We do not
consider here mechanisms leading to disclosure of information and
its consequences. It seems reasonable that the measure
$S(p_{0},\dots ,p_{N})$ we would like to use for measuring the
investor behaviour (or more precisely the lottery defined by the
probabilities $p_{i}$) be additive in the following sense. It
should not matter if the portfolio is constructed in one draw with
the probabilities ${p_{0}, p_{1}, p_{2}, \dots , p_{N}}$ or with
two subsequent draws with probabilities $\frac{p_{0}}{p_{0}+p_{1}}
,\frac{p_{1}}{p_{0}+p_{1}}$  and $p_0+p_1,p_{2}, \dots , p_{N}$.
This leads to the equation: $$\begin{array}{rcl} S\left(
p_{0},p_{1}, \dots ,p_{N}\right)& = &S\left( p_{0}+p_{1}, p_{2},
\dots ,p_{N} \right) +\\&&+ \left( p_{0}+p_{1}\right) S\left (
\frac{p_{0}}{p_{0}+p_{1}},\frac{p_{1}}{p_{0}+p_{1}}\right) ,
\end{array}\eqno(4)$$ where the arguments describe probabilities
of drawing (that is buying) of the assets. We will look for a
solution fulfilling: $$S\left( p_{0},p_{1}, \dots ,p_{N}\right) =
S\left( p_{0}\right) + S\left( p_{1}, p_{2}, \dots ,p_{N} \right)
\eqno(5)$$

Equations $(4)$ and $(5)$ after some standard algebraic
manipulations lead to the following solution for the function $S$,
see \cite{10} : $$S\left( p_{0},p_{1}, \dots ,p_{N}\right) = -\sum
_{k=0}^{N} p_{k} \ln p_{k} \ . \eqno(6)$$ For obvious reasons, we
will call this function entropy.  The number $K:=e^S$ gives the
effective number of assets in the portfolio: $K$ is equal $M$ if
the weights are distributed uniformly between $M$ assets, and is
$1$ if the portfolio contain a single asset.\\

Let us now consider the "cartesian product" of two statistically
independent portfolios consisting of $N+1$ and $M+1$ assets,
respectively. Then
$$\begin{array}{l}S_{\left( M+1\right)\times \left( N+1\right)} =
- \sum _{k=0}^{M} \sum _{l=0}^{N} p_{k} p'_{l} \ln \left( p_{k}
p'_{l}\right) =\vspace{2ex}\\= -\sum _{k=0}^{N} p_{k} \ln \left(
p_{k}\right) -\sum _{l=0}^{N} p'_{l} \ln \left( p'_{l}\right) =
S_{M+1} + S_{N+1}\end{array} \eqno(7)$$ so the entropy is
additive. We would like to compare returns ("achievements")  of
portfolio managers (owners). Therefore we classify the investors
according to the value of their portfolio. This would allows us to
divide the appropriate returns into two parts corresponding to
manager's lore and tide of the market. We will choose the
portfolios that maximize the entropy to represent the classes of
investors. An "external" observer would know only the expectation
value of returns. (We suppose that draws are not correlated and
the market is efficient.) Perhaps, it would be better to assume
that the values of the random variable $\kappa $ correspond to
whole sectors of the markets (e.g. oil companies) rather then to
separate assets \cite{1, 11}. This means that we are looking for a
conditional extrema of the function $S(p_{0}, \dots ,p_{N})$. The
conditions are: $$\sum _{k=0}^{N} p_{k}=1 \eqno(8a)$$ and $$
-c_{00}\left( c_{0},\dots ,c_{N} \right)= E\left( \sum _{k=0}^{N}
\left[ \kappa = k\right] c_{k} w_{k} \right) \eqno(8b)$$ We have
weakened the balance condition (2) to be fulfilled only on the
expectation values level. The Lagrange method of finding
conditional extrema requires the following differential form to
vanish: $$dS\left( p_{0}, \dots, p_{N}\right) + \beta dE\left(
\sum _{k=0}^{N} \left[ \kappa = k\right] c_{k} w_{k}\right) +
\gamma d\sum _{k=0}^{N} p_{k}=0 , \eqno(9) $$ where $\beta$ and
$\gamma $ are Lagrange multipliers. The substitution of equation
(6) leads to the condition $$ -\ln p_{k} -1 +\beta c_{k} w_{k} +
\gamma =0 \eqno(10)$$ which gives explicit dependence of the
probabilities $p_{k}$ characterizing the maximal entropy portfolio
on the prices $c_{k}$: $$ p_{k}= \exp \left( \beta c_{k}w_{k}
+\gamma+1\right)\eqno(11)$$ The multiplier $\gamma$ can be
eliminated by explicit normalization and the elimination leads to
Gibbs-like probability distribution: $$p_{k}\left( c_{0},\dots
,c_{N}\right) = \frac{  \exp \left( \beta c_{k}w_{k} \right) }
{\sum _{k=0}^{N}\exp \left(  \beta c_{k}w_{k} \right) }.
\eqno(12)$$ Now we are in a position to define {\it canonical
statistical ensemble} of investors which consists of all investors
that have got the same return. The canonical ensemble describes
all strategies (zombies and gamblers) leading to the same return.
It can be represented by the portfolio maximizing the entropy
({\sl canonical portfolio}). Choosing any other representative
would mean lower entropy and as a result would give a bias to one
strategy (knowledge). If one recalls that $K=\exp S$ gives the
effective number of assets in the portfolio then one immediately
gets that the entropy $S$ measures also the diversification of the
portfolios. The canonical portfolio is the safest one in the class
of the same return. We will also define the temperature $T$ of the
canonical ensemble as $T:=\frac{1}{\beta}$ and the statistical sum
$Z$ as $Z\left(c_{0}, \dots ,c_{N}\right) := \sum _{k=0}^{N} \exp
( \beta c_{k}w_{k} ) $. If we keep the weights $w_{k}$ constant
then the changes of the prices $c_{k}$ imply appropriate changes
of the value of the portfolio $-c_{00}$ (return). The expected
infinitesimal change of $c_{00}$ is given by $$-dc_{00}\left( c_{0},
\dots , c_{N}\right) = dE\left( c_{\kappa} w_{\kappa}\right)
.\eqno(13) $$ Having in mind that $S=-\sum_{k=0}^N p_{k}\ln
p_k=\ln Z - \beta E\left( c_{\kappa}w_{\kappa} \right) $ we can
write
$$\begin{array}{rcc}dS&=&\frac{\beta \sum_{k=0}^N w_{k} \exp
\left( \beta w_{k}c_{k}\right) dc_{k}}{Z\vphantom{^{2^2}}}- \beta
dE\left( c_{k}w_{k}\right)~~ =\vspace{2ex}\\&=& \beta \left(
\sum_{k=0}^N E \left( \left[ \kappa =k\right] w_{k} \right)
dc_{k}-dE\left( c_{\kappa}w_{\kappa}\right)\right) .\end{array}
\eqno(14)$$ This implies that if we treat $S$ as an independent
variable $$dc_{00}\left( c_{0},\dots , c_{N}, S\right)+ \sum
_{k=0}^{N} \overline{w}_k dc_{k} = TdS , \eqno(15)$$ where
$\overline{w}_k$ denotes the mean content of the $k$-th asset in
the portfolio. Now, the temperature of the portfolio $T$ is equal
to
$$ T=\frac{\partial c_{00}}{\partial S}. \eqno(16)$$ This
temperature measures the change of the portfolio value caused by
its entropy change. As in classical thermodynamics, we can
formulate two principles. \\ {\bf The I principle of the canonical
ensemble}: {\em The change of value of a canonical portfolio
$-c_{00}$ consists of two parts. The first one is equal to the
change the investors knowledge $\delta Q$ and the second is equal
to change of the values of content of the portfolio:} $$ dc_{00}
+\delta Q +\sum_{k=0}^N \overline{w}_k dc_{k}=0. \eqno(17)$$ {\bf
The II principle of the canonical ensemble}: {\em The value of
lost investors knowledge $-\delta Q$ is proportional to the
increase in the entropy of its canonical ensemble:} $$ \delta Q +
TdS =0 . \eqno(18)$$ Note that we have written $\delta Q$ instead
of the obvious $dQ$ because, in general, such a function $Q$ does
not exist. Following the development of  thermodynamics we will
define the {\it free value} $-F$ of the canonical portfolio as: $$
F\left( c_{0},\dots , c_{N},T\right) :=c_{00}\left( c_{0},\dots
,c_{N},S\right)-TS .\eqno(19)$$ This allows for the formulation of
the two principles of the canonical ensemble in a balance-like
equation:
$$dF\left( c_{0},\dots , c_{N},T\right) +SdT + \sum _{k=0}^{N}
\overline{w}_{k}dc_{k} =0 \eqno(20)$$ which combined with the
definition of the entropy $S$ gives $$-F=T\ln Z \eqno(21a)$$ and
$$-c_{00}=\frac{\partial \ln Z}{\partial \beta}.\eqno(21b)$$ The
name free value can be justified as follows. Let us suppose that
during the market evolution (changes of the prices $c_{k}$) the
class of the investor does not change ($T=const$). Then the
changes of values of the assets in the portfolio, $\sum_{k=0}^N
\overline{w}_{k}dc_{k}$ are measured by changes of the potential
$F$. Such a process can be referred to as isothermal. $-T$ can be
interpreted as price of a unit of the entropy $S$. The entropy of
a canonical portfolio increases when the absolute value of the
parameter $T$ increases. This means that an amateur pays more than
a professional investor for errors of the same order. But if the
temperature is negative an amateur gets more from erroneous
decisions. A portfolio with $T<0$ and small entropy can only be
constructed by a specialist who uses his knowledge in reverse.
After changing signs of the weights $w_{k}$ (short position) such
a portfolio gives excellent returns. Note that due to the
additivity of entropy the temperature of the portfolio constructed
by merging  two portfolios with temperature $T$ but built in
different market sectors equals $T$. \\

It would be helpful to give some flesh to the above consideration
by working out a numerical example. Let us consider two investors
Alice and Bob. Both have the same initial capital, say \$1 and,
besides money, there are only two assets available on the market,
$a_{1}$ and $a_{2}$. Bob (a zombie) divided his capital into two
equal parts and spent them on both assets. Alice (a zombie, a
pro?) spent a quarter of her capital on $a_{1}$ and kept the rest.
Suppose now that after some time the price of the asset $a_{1}$
went down by 20\% and the price of the asset $a_{2}$ increased by
30\% that is the relative prices are now $c_{1}=0.8$ and
$c_{2}=1.3$. This means that Alice's and Bob's portfolios gave the
returns $c_{00}^{A}=0.95$ and $c_{00}^{B}=1.05$, respectively (we
have neglected the interest rates). The temperatures are
$T^{A}=-0.46$ and $T^{B}=2.55$, respectively (only two decimal
positions are kept). Bob has got better return then Alice even
though $T^{B}>T^{A}$. This is because Bob alone makes profit from
his knowledge ($T^{B}$ is positive). But the authors would prefer
Alice to Bob as an investment adviser because when one listen to
her advises and acts contradictory to them one gets better return
then Bob's. Alice's knowledge is greater than Bob's because if
$\mid T^{B}\mid > \mid T^{A}\mid $ then $S^{B} > S^{A}$.

\section{Conclusions}
We have proposed a method that allows numerically measure
investors qualities. Inspired by thermodynamics, we were able to
define canonical ensembles  of portfolios, the temperature of
portfolios and, possibly, various thermodynamics-like potentials.
We have used the relative prices $c_{k}$ of the assets that have
direct interpretation. The theory of financial market would prefer
a covariant description of the portfolio. This can be easily
achieved by replacing the parameters $c_{00}, c_{1}, \dots ,c_{N}$
by their natural logarithms (that is integrals of the
instantaneous interest rates):

$$\ln \mid c_{00}\mid\thinspace \rightarrow c_{00} \eqno(22a)$$

$$\ln \mid c_{k}\mid\thinspace \rightarrow c_{k} \eqno(22b)$$

$$ \frac{\mid c_{k}\mid}{\mid c_{00} \mid}\thinspace w_{k}
\rightarrow w_{k}. \eqno(22c)$$

We imagine that analogous methods can be developed for valuation
of credit repayment scenarios or bonds under the stochastic
behaviour of interest rates.

                                                                                                                                                                                                                                                                                                                                                                                                                                                                                                                                                                                                                                                                                                                                                                                                                                                                                                                                                                                                          !
                                 
\end{document}